\documentclass[english]{article}
\usepackage[T1]{fontenc}
\usepackage[latin9]{inputenc}
\usepackage{geometry}
\usepackage{babel}
\usepackage{textcomp}
\usepackage{url}
\usepackage{amstext}
\usepackage{setspace}
\makeatletter
\newcommand{\lyxaddress}[1]{
\par {\raggedright #1
\vspace{1.4em}
\noindent\par}
}

\usepackage{fixltx2e}
\usepackage{booktabs}
\usepackage{tabularx}
\usepackage{pgfplots}
\usepackage{graphicx}
\date{}

\makeatother

\usepackage{listings}

\begin{document}

\title{A Performance Survey on Stack-based and Register-based Virtual Machines}

\author{Ruijie Fang, Siqi Liu}
\maketitle

\lyxaddress{\begin{center}
ruijie.fang@temple.edu, tylerliu2018@lschs.org 
\par\end{center}}
\begin{abstract}
Virtual machines have been widely adapted for high-level programming
language implementations and for providing a degree of platform neutrality.
As the overall use and adaptation of virtual machines grow, the overall
performance of virtual machines has become a widely-discussed topic.
In this paper, we present a survey on the performance differences
of the two most widely adapted types of virtual machines - the stack-based
virtual machine and the register-based virtual machine - using various
benchmark programs. Additionally, we adopted a new approach of measuring
performance by measuring the overall dispatch time, amount of dispatches,
fetch time, and execution time while running benchmarks on custom-implemented,
lightweight virtual machines. Finally, we present two lightweight,
custom-designed, Turing-equivalent virtual machines that are specifically
designed in benchmarking virtual machine performance - the \textquotedblleft Conceptum\textquotedblright{}
stack-based virtual machine, and the \textquotedblleft Inertia\textquotedblright{}
register-based virtual machine. Our result showed that while on average
the register machine spends 20.39\% less time in executing benchmarks
than the stack machine, the stack-based virtual machine is still faster
than the virtual machine regarding the instruction fetch time.
\end{abstract}
\begin{description}
\item [{Keywords:}] Virtual Machine, Interpreters, Stack Architecture,
Register Architecture
\end{description}

\section{Introduction}

Virtual machines are both ideal and popular for high-level and very-high-level
interpreted programming language implementations. Virtual machines
add a layer of abstraction above the lower-level hardware resources,
providing more flexibility for language implementers to implement
advanced functionalities while bringing the benefits of being architecture-neural
and portable. A bytecode interpreter is a core part of a virtual machine,
providing the capacity to read and evaluate instructions in an intermediate
bytecode-format. Most virtual machines are implemented on either the
stack-based architecture, employing stacks as a means of data storage
and manipulation, or the register-based architecture, simulating the
architecture of a physical computer.

A widely discussed question in the field of virtual machine architecture
is that whether the stack-based architecture or the register-based
architecture has a better performance. The stack-based architecture,
being widely used, is known for its ease of implementation, as the
location of operands does not need to be explicitly specified \cite{Davis:2003:CVR:858570.858575,Ertl:1995:SCI:207110.207165,ertl:1996}.
In contrast, the register architecture requires the location of operands
to be specified manually. The difference in specifying the location
of the operands can be better recognized through the typical difference
in instruction statements, as in stack-based virtual machines, usually
only one parameter is supplied for a specific instruction (the pseudo-bytecode
below presented in a format similar to the JVM):

\begin{lstlisting}
iconst 1
iconst 2
iadd
\end{lstlisting}

In contrast, in a register-based virtual machine's bytecode format,
more parameters specifying the address of the operands are required
(the pseudo-bytecode below presented in a format similar to the Parrot
VM):

\begin{lstlisting}
set t3, t1
set t4, t2
add t3, t4, t5
\end{lstlisting}

The needlessness to specify the operand's address created a huge benefit
for the stack-based architecture, as the bytecode format can thus
be simpler, simplifying the overall code generation process. The overall
code size needed to construct a stack-based virtual machine is also
smaller, as compared to the register-based architecture. The stack-based
architecture has been widely adapted, most notably on the Java Virtual
Machine (JVM) for running Java bytecodes, and on CPython for interpreting
Python source codes.

However, the register architecture still nonetheless contains a notable
amount of decent features. The explicit definition of address of the
operands would allow the bytecode size to be reduced, which overall
contributes to the specific virtual machine's performance \cite{Shi:2005:VMS:1064979.1065001}.
Moreover, the application of the register-based architecture on the
newer versions of Lua \cite{lerusalimschy:2005} and the Parrot VM
\cite{fagerholm:2005} suggests that in many cases, well-written register
virtual machines could outperform their stack-based rivals in terms
of performance and execution time.

Much research related to the performance comparison between the stack
architecture and the register architecture had been done, most notably
the two papers focusing on comparing the original, stack-based JVM
(the Java Virtual Machine) to a custom-translated, register based
version of the JVM by Davis et. al and Shi et. al \cite{Davis:2003:CVR:858570.858575,Shi:2005:VMS:1064979.1065001},
where the result indicated that the register bytecode required fewer
instructions than a stack-based architecture's bytecode. However,
most research papers addressing this topic focused on either comparing
the performance on existing, general purpose virtual machine platforms,
or translating an existing stack-based virtual machine into a register-based
VM. In this paper, we present a new approach towards comparing stack-based
architecture and register-based architecture by constructing two domain-specific
virtual machines made specially for the purpose of benchmarking and
comparison - the Conceptum stack-based virtual machine and the Inertia
register-based virtual machine. Under such a setting, both the simplicity
and the similarity in the overall structure and format of both architectures
are guaranteed, thus leading to better and more accurate results.
Moreover, we applied a more straightforward method in evaluating the
overall performance of virtual machines by measuring the time spent
in instruction fetch, instruction dispatch, execution and the total
amount of dispatches needed. In the following sections, we provide
a brief overview of the structure and construction of Conceptum and
Inertia and the benchmark results on the two separate platforms. Finally,
we present an overall analysis of each of the architecture's performance.

\section{The Conceptum stack-based virtual machine}

Conceptum is a stack-based virtual machine specially designed for
performance benchmarks. Conceptum is written in pure, ANSI C11-compliant
C. Conceptum is type-safe, Turing-equivalent, and utilizes a switch
as an instruction dispatch method. Our reason for using switch dispatch
(for a deeper discussion, see section 4.1) is that although switch
dispatch might not be the fastest, it is still widely adopted \cite{NewZakeski}.
Inertia - the register-based virtual machine - had adopted the same
dispatch method.

In terms of computability, Conceptum is Turing-equivalent, as Conceptum's
design resembles that of a 2x pushdown automata (2-PDA). It had been
proven that a 2-PDA is Turing-equivalent, as it employs two stacks
to achieve the ability of random and infinite memory access \cite{Koslowski:2013}.
Conceptum always preserves two stack spaces for the bytecode program
to operate on: one local-scope stack for procedure execution, garbage-collected
right after the procedure returns, and another global stack for exchanging
arguments and return results between procedure calls and preserving
data.

Conceptum is designed to act like a Just-In-Time interpreter. Once
Conceptum's assembler reads in a program, the assembler parses the
program, breaking the procedures into parsed bytecodes, and feeding
them into Conceptum's run() function. The run() function initializes
two stacks of fixed size, defined by a global macro. One stack is
initialized as a global stack, whereas another stack initialized as
the local stack for the main bytecode procedure. The run() function
then feeds the parsed bytecode and the reference of the two initialized
stacks into the eval() function where the switch dispatch exists,
and the eval() statements loops until the program comes to a halt.
A bytecode piece in Conceptum is defined as a struct:

\begin{lstlisting}
 typedef struct {
    int32_t instr;
    void *payload;
 } ConceptInstruction_t;
\end{lstlisting}

The procedure itself is represented as an array of such structures,
whereas the program is represented as an array of an array of procedures,
where the first procedure is being recognized as the main procedure.
Given a struct, the switch inside eval() compares the integer value
`instr` inside the struct to a set of previously-defined macros to
dispatch the instruction.

Procedure calls in Conceptum are handled at the bytecode compilation
stage by the Conceptum assembler. Iterating through every line of
the unparsed source file, the assembler stores every procedure's name
into another array of the same order as the program itself (which
is in turn, an array of procedures, as we previously stated). The
assembler then utilizes a simple lexical analyzer to parse each of
the procedures line by line. When a call statement occurs, the lexical
analyzer performs an O(n) search inside the array of procedure names;
when a match is found, the analyzer substitutes in the address (which
is simply the array index) as the calling address in the parsed bytecode's
payload. This mechanism is invented in order to reduce the procedure
call cost in the actual instruction dispatch; by substituting in the
address of a procedure, the complexity of a procedure call in the
actual dispatch will always be O(1). Note that in order for this described
procedure name substitution to work, the main procedure where Conceptum
starts off calling needs to be coded on the very top of the program
\cite{fang:201608}.

Conceptum's eval() function containing the switch dispatch functions
is based on four variables: two stacks (defined in the caller function
and passed in as variables), one instruction array (which is an array
of procedure arrays, as previously noted), and an integer index indicating
which procedure inside the program array to execute. Immediately after
reading in the four parameters, a for loop iterates over the designated
procedure to execute, supplying every single line of instructions
to the switch for dispatch. In the actual switch dispatch, a procedure
call in the bytecode would make the eval() function call upon itself,
but simply supplying another integer index (the index already substituted
in as an operand by the assembler, as previously described). In such
calls, the global stack is preserved while a new, local stack for
the procedure being called must be allocated. The eval() function
is defined to return a void pointer:

\begin{lstlisting}
void* eval(int index,
  ConceptStack_t *stack, ConceptStack_t *global_stack,
  int start_by);
\end{lstlisting}

The void pointer being the first element in the local stack of a procedure,
therefore allowing procedures to return values. The benefit of eval()
calling itself when encountering a new procedure call is that then
the recursive procedure calls inside the bytecode could be supported,
as recursive procedure calls simply indicate recursions in the eval()
function with the creation of new local stacks. While the overall
design of Conceptum can be viewed as that of a 2-PDA's in order to
prove that Conceptum is Turing-equivalent, the support for recursion
also can be used to indicate Conceptum's Turing-equivalence. For simplicity
and speed, procedure calls do not directly take arguments in Conceptum.
Instead, while preserving a unique local stack, all procedures have
access to the one, single global stack. To supply arguments to the
procedure call, programs can push values into the global stack, which
can be accessed later by the procedure being called upon.

In terms of garbage collection and memory management, Conceptum behaves
similar to the JVM. All values declared within the local scope of
the procedure will immediately be garbage collected (by iteratively
freeing every single void pointer on the local stack) after the procedure's
return, and everything in the global stack gets garbage collected
once the main procedure returns. It can be said that the garbage collection
process is rather simple in Conceptum; after all, every value is stored
on the stack, which freeing the garbage collector would simply free
up the stacks at appropriate times. While there exists no functions
specifically made for garbage collection purposes in the source code
for Conceptum, a custom wrapper library around C's dynamic memory
allocation functions that registers every use of malloc() in a stack
does exist, in order to prevent memory leaks \cite{fang:2016160466}.

\section{The Inertia register-based virtual machine}

Inertia is a register-based virtual machine specifically designed
in order to resemble Conceptum in its complexity. For benchmark results
to be accurate, the overall similarity of the structure of both Conceptum
and Inertia needs to be ensured. In the paragraphs below, we present
a brief overview of Inertia's design. By default, the Inertia virtual
machine reads in bytecodes in binary format. Inertia's counterpart,
the assembler ``Moment of Inertia'', is the syntax-based assembler
that reads in programs written in text-based instructions and translates
them into bytecode \cite{liu:2016888}. It is necessary to state here,
that neither of the time spent on file I/O and the time spent on decoding
and parsing the program being read in is measured during the actual
benchmark, except the actual execution time, after the bytecodes are
being parsed into internal data structures.

In order to start an execution cycle, Inertia's main() function reads
in the bytecode, feeds the bytecodes into program array, and starts
the execution of the program by run(). The run() function initializes
the program counter and a flag variable, used to determine whether
if the program is still running. The run() function follows the fetch-execution
cycle, executing parsed bytecode programs by calling the decode()
and eval() functions. The decode() function fetches the program instruction
by calling fetch(). The bytecode piece in Inertia, similar to that
in Conceptum, is defined as a struct:

\begin{lstlisting}
typedef struct {
	uint32_t instr;
} I_instr_t 
\end{lstlisting}

Although one might notice that it seems wasteful to wrap the variable
definition 

\begin{lstlisting}
uint32_t instr; 
\end{lstlisting}

around a struct, the wrapping is intended for both the overall fairness
of the benchmark and the consistency of style over Conceptum and Inertia,
as Conceptum also uses a struct to represent an instruction: 

\begin{lstlisting}
typedef struct {
    int32_t instr;
    void *payload;
 } ConceptInstruction_t;
\end{lstlisting}

In order to be consistent with Conceptum, Inertia uses a similar switch
dispatch method. A loop is used to iterate over instructions, which
every instruction gets fed into the eval() function containing the
large switch statement. In order to resemble Conceptum\textquoteright s
switch dispatch process, all operations were not executed in the switch
statement, but contained in separate, external function definitions.
Garbage collection in Inertia also resembles that of Conceptum. On
the start, a fixed-size memory space for data storage is allocated.
The space is only freed after the virtual machine halts. 

Inertia is also similar to Conceptum in terms of complexity of procedure
call instructions. Both Conceptum and Inertia has an O(1) complexity
in procedure calls. To achieve the O(1) complexity, Inertia does not
preserve a local memory space for procedures. All procedures operate
in the same, global, memory space. Since only a global memory space
is required, recursion is supported in Inertia. When encountering
a recursive procedure call, Inertia\textquoteright s eval() function
simply evokes a new run() function, which starts at the beginning
of the specific procedure.

\section{Evaluating performance}

\subsection{Performance measurement}

In this section, we present our method of evaluating the performance
of Conceptum and Inertia. It is commonly agreed upon in previous research\cite{Davis:2003:CVR:858570.858575,Shi:2005:VMS:1064979.1065001,Myers:1977:CAS:859402.859403,mcglashan:1999,Schulthess:1977:RCA:859412.859416,winterbottom:1997}
that register machines outperform stack machines in both bytecode
size and execution performance. However, few results exist in using
the measurement of the actual running time of benchmarks to evaluate
a virtual machine's performance. \cite{Davis:2003:CVR:858570.858575}
proposed an estimation formula for estimating the performance differences
between a stack-based machine (VSM) and a register-based machine (VRM):

\[
T
\textsubscript{VRM}
\ensuremath{\approx}
T\textsubscript{VSM}
\text{\textminus}\#dispatches
\text{\texttimes}
T
\textsubscript{dispatch}
+\#fetches
\text{\texttimes}
T
\textsubscript{fetch}
\]

The equation above states that the correlation between the running
time of the register-based and the stack-based virtual machine is
mainly correlated to the difference in the amount of dispatches, the
cost of a dispatch, the cost of instruction fetches and the difference
in the number of instruction fetches. Analyzing the equation, one
can see that the register bytecode contains more instruction fetches
than the bytecode of a stack-based virtual machine, as the stack-based
virtual machine contains only an optional operand per instruction;
in contrast, the number of operands in a register machine can vary
from 1 to 4 (as in the design of Inertia). However, one could also
see that the cost of such increase in the overall amount of fetches
to be relatively small, as one fetch would simply imply the action
of loading one more operand from an operands array. The difference
in the amount of dispatches and dispatch time is also an important
part of \cite{Davis:2003:CVR:858570.858575}'s equation. As the overall
amount of instructions in a bytecode program for the register machine
is generally regarded as shorter than that of a program for the stack-based
machine, having shorter dispatch time might be a significant advantage
of register machines. However, \cite{Davis:2003:CVR:858570.858575}
and a few other research also noted that the switch dispatch method
(used in both Conceptum and Inertia) is not the most efficient dispatch
method - probably even the most inefficient one, as compared to threaded
dispatch or direct threaded dispatch. Our view towards this problem
is that, since switch dispatch is still commonly applied among virtual
machines \cite{NewZakeski}, and that both Conceptum and Inertia adopted
the same switch dispatch method in instruction dispatch, the inefficiency
of switch dispatch is not likely to impact the benchmark results.

Based on the equation provided by \cite{Davis:2003:CVR:858570.858575}
and information from other research \cite{Shi:2005:VMS:1064979.1065001,Myers:1977:CAS:859402.859403,Schulthess:1977:RCA:859412.859416},
we propose a method in measuring the performance of Conceptum and
Inertia. Under such method, the overall performance will be evaluated
using four different measurements, three of which are based on timing
intervals of the execution time and one measuring the amount of dispatch.
The three different timing intervals are: the overall bytecode execution
time, timed right after the bytecode file has been read and parsed,
and ending after every single instruction finished executing; the
accumulative dispatch time, as a sum of the time spent for every single
switch dispatches. The lesser the time is, the better the performance
will be. It is also important to note that the overall amount of dispatches
is also measured to show the correlation between the overall dispatch
time and the total amount of dispatches. Note that the execution time
does not include the time of reading in the program from bytecode
or parsing the program. Since the syntax for the bytecode of a register-based
virtual machine and the syntax for that of a stack-based virtual machine
are inherently different, Conceptum's syntax-based assembler and Inertia's
syntax-based assembler were programmed in different fashions and cannot
be equally compared. Moreover, the time for the file I/O is generally
nondeterministic and fluctuates based on other I/O intensive programs
being executed on the same host system, accounting file I/O would
only add a layer of inaccuracy to the overall benchmark results.

\subsection{A note on timing}

To more accurately time the execution, we measured the CPU time, instead
of the wall clock time, for the benchmarks. Measuring CPU time is
a common method applied in real-time systems, in order to achieve
the most accurate timing. The CPU time in GNU C refers to the time
a process has used the CPU since the start of the process \cite{NewGnuClockTime}.
Therefore, by using the CPU time, minor disturbances usually present
in the wall clock time such as process preemptions can be reduced.
In standard C, the current CPU time in a process is given by the function
\emph{clock()}, which returns a \emph{clock\_t }type, indicating the
total clock ticks spent in computing since the start of the process
{[}20{]}. To present the timing results in an easily readable format,
we utilize the code below to convert a the CPU time given by the clock()
function to microseconds:

\begin{lstlisting}
clock_t time_in_microseconds = clock() * 1000000 / CLOCKS_PER_SEC;
\end{lstlisting}

where \emph{CLOCKS\_PER\_SEC} is a macro provided in the standard
C library header\emph{ time.h} .

Based on our algorithm described above, the total time spent by both
Conceptum and Inertia executing a program is measured as:

\begin{lstlisting}
clock_t begin_execution_time = clock();
run(); // execute the parsed bytecodes already read in
clock_t end_execution_time = clock();
clock_t time_in_microseconds = (end_execution_time - begin_execution_time);
\end{lstlisting}

However, given that every single \emph{clock()} is a function call
and produces at minimum a function call overhead, simply applying
the method of measuring the CPU time at the start of an interval,
measuring the CPU time at the end of an interval, and subtracting
the start CPU time from the end CPU time to get the total time spent
in an interval isn't enough. To provide even more accurate measurements,
we adopted the method described in \cite{NewBaker:2016}, where the
time spent calling the clock() function is first measured, then subtracted
from the overall execution time:

\begin{lstlisting}
t0 = clock();
t1 = clock();
run();
t2 = clock();
d = (t2 - t1) - (t1 - t0);
\end{lstlisting}

Finally, it is imperative to note that since both dispatch time and
fetch time are measured, and that both the dispatch time and fetch
time are contained in the execution time measured, the overall execution
time might not be the most accurate, as the final execution time combines
the overheads produced by the extra measurement of fetch and dispatch
time. However, since such inaccuracy is contained in all benchmarks,
it shouldn't affect the final result.

\section{Performance benchmarks}

We utilize the results of multiple benchmark programs written in bytecode
format to compare the performance differences in the register architecture
and the stack architecture. As stated in the pervious section, the
overall time of bytecode execution (starting after the bytecodes were
read in and parsed, and ending after the last instruction finished
its execution), the average dispatch time, and the overall dispatch
time (sum of the time of every single dispatch) and the total amount
of dispatches executed serve as the four benchmark results evaluating
the overall performance.

All benchmark programs are executed on Conceptum and Inertia, where
both virtual machines were deployed on a physical test machine with
an Intel Core i5 processor with 8 Gigabytes of RAM. The compiler used
to compile both Conceptum and Inertia is gcc 4.9.4. Both Conceptum
and Inertia are compiled without any compiler optimizations, with
the compiler flags:

\begin{lstlisting}
gcc -O0 -std=c11
\end{lstlisting}

To show the most accurate performance of both machines, a few benchmarks
had been selected, featuring either intensive read/write operations
on the stacks or registers or an intensive amount of arithmetic instructions
which benchmarks the overall dispatch performance. The benchmarks
we have selected are:
\begin{itemize}
\item Fibonacci: A program that calculates Fibonacci numbers up to 1,000
times, and stores them in the memory.
\item ExhaustiveCollatz: A program that proves the Collatz conjecture for
numbers below 20,000.
\item AddictiveAddition: A program that increments from 0 to $5*10^{6}$,
writing every result into the memory.
\item Recursion: A program that contains a procedure that recursively calls
itself 1,000 times, while decreasing an integer counter.
\end{itemize}
All times are measured in microseconds, while all benchmarks have
ran multiple times, the final result being the average of every single
execution. The results are presented below.

\subsection{Total amount of dispatches}
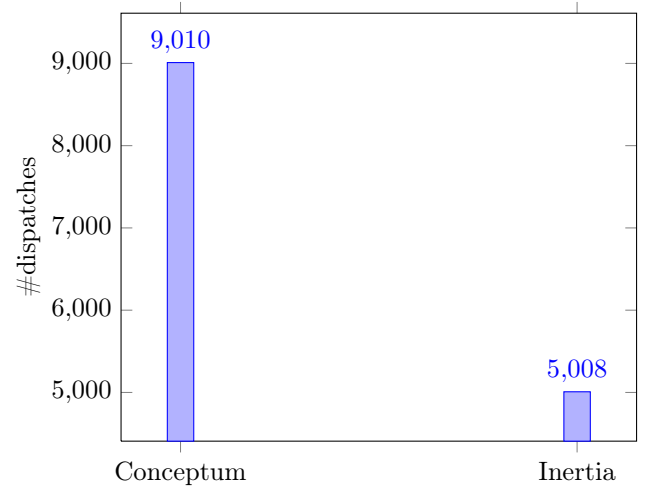
\begin{center}
\noindent\begin{minipage}[t]{1\columnwidth}%
\begin{minipage}[t]{0.45\columnwidth}%
\begin{tikzpicture}
\begin{axis}[
  title=Figure 5.1-A: Fibonacci,
  ybar,
  enlargelimits=0.15,
  legend style={at={(0.5,-0.15)},
  anchor=north,legend columns=-1},
  ylabel={\#dispatches},
  symbolic x coords={Conceptum,Inertia},
  xtick=data, nodes near coords,
  nodes near coords align={vertical}, ]
\addplot coordinates {(Conceptum,14012) (Inertia,3008)};
\end{axis}
\end{tikzpicture}%
\end{minipage}\hfill{}%
\begin{minipage}[t]{0.45\columnwidth}%
\begin{tikzpicture}
\begin{axis}[
  title=Figure 5.1-B: ExhaustiveCollatz,
  ybar,
  enlargelimits=0.15,
  legend style={at={(0.5,-0.15)},
  anchor=north,legend columns=-1},
  ylabel={\#dispatches},
  symbolic x coords={Conceptum,Inertia},
  xtick=data, nodes near coords,
  nodes near coords align={vertical}, ]
\addplot coordinates {(Conceptum,30850935) (Inertia,19114675)};
\end{axis}
\end{tikzpicture}%
\end{minipage}%
\end{minipage}
\par\end{center}

\noindent \begin{center}
\noindent\begin{minipage}[t]{1\columnwidth}%
\begin{minipage}[t]{0.45\columnwidth}%
\begin{tikzpicture}
\begin{axis}[
  title=Figure 5.1-C: AddictiveAddition,
  ybar,
  enlargelimits=0.15,
  legend style={at={(0.5,-0.15)},
  anchor=north,legend columns=-1},
  ylabel={\#dispatches},
  symbolic x coords={Conceptum,Inertia},
  xtick=data, nodes near coords,
  nodes near coords align={vertical}, ]
\addplot coordinates {(Conceptum,35000007) (Inertia,20000005)};
\end{axis}
\end{tikzpicture}%
\end{minipage}\hfill{}%
\begin{minipage}[t]{0.45\columnwidth}%
\begin{tikzpicture}
\begin{axis}[
  title=Figure 5.1-D: Recursion,
  ybar,
  enlargelimits=0.15,
  legend style={at={(0.5,-0.15)},
  anchor=north,legend columns=-1},
  ylabel={\#dispatches},
  symbolic x coords={Conceptum,Inertia},
  xtick=data, nodes near coords,
  nodes near coords align={vertical}, ]
\addplot coordinates {(Conceptum,9010) (Inertia,5008)};
\end{axis}
\end{tikzpicture}%
\end{minipage}%
\end{minipage}
\par\end{center}

\subsection{Total time of instruction fetches, total time of instruction dispatches
and total execution time}
\begin{center}
\noindent\begin{minipage}[t]{1\columnwidth}%
\begin{minipage}[t]{0.45\columnwidth}%
\begin{tikzpicture}
\begin{axis}[
title=Figure 5.2-A: Fibonacci,
ybar,
enlargelimits=0.15,
legend style={at={(0.5,-0.15)},
anchor=north,legend columns=-1},
ylabel={\#time (microseconds)},
symbolic x coords={Fetch time,Dispatch time,Execution time},
xtick=data,
nodes near coords, nodes near coords align={vertical}, ]
\addplot coordinates {(Fetch time,6126.867) (Dispatch time,7319.733) (Execution time,25711)};
\addplot coordinates {(Fetch time,2690.133) (Dispatch time,1339.067) (Execution time,8058.4)};
\legend{Conceptum,Inertia} \end{axis} \end{tikzpicture}%
\end{minipage}\hfill{}%
\begin{minipage}[t]{0.45\columnwidth}%
\begin{tikzpicture}
\begin{axis}[
title=Figure 5.2-B: ExhaustiveCollatz,
ybar,
enlargelimits=0.15,
legend style={at={(0.5,-0.15)},
anchor=north,legend columns=-1},
ylabel={\#time (microseconds)},
symbolic x coords={Fetch time,Dispatch time,Execution time},
xtick=data,
nodes near coords, nodes near coords align={vertical}, ]
\addplot coordinates {(Fetch time,11887959.87) (Dispatch time,14282572.93) (Execution time,50052119.67)};
\addplot coordinates {(Fetch time,19784777.33) (Dispatch time,7802168.267) (Execution time,55919524.4)};
\legend{Conceptum,Inertia} \end{axis} \end{tikzpicture}%
\end{minipage}

\begin{minipage}[t]{0.45\columnwidth}%
\begin{tikzpicture}
\begin{axis}[
title=Figure 5.2-C: AddictiveAddition,
ybar,
enlargelimits=0.15,
legend style={at={(0.5,-0.15)},
anchor=north,legend columns=-1},
ylabel={\#time (microseconds)},
symbolic x coords={Fetch time,Dispatch time,Execution time},
xtick=data,
nodes near coords, nodes near coords align={vertical}, ]
\addplot coordinates {(Fetch time,11641908.53) (Dispatch time,13959241.13) (Execution time,48963810)};
\addplot coordinates {(Fetch time,19557117.27) (Dispatch time,8425430.067) (Execution time,56730777)};
\legend{Conceptum,Inertia} \end{axis} \end{tikzpicture}%
\end{minipage}\hfill{}%
\begin{minipage}[t]{0.45\columnwidth}%
\begin{tikzpicture}
\begin{axis}[
title=Figure 5.2-D: Recursion,
ybar,
enlargelimits=0.15,
legend style={at={(0.5,-0.15)},
anchor=north,legend columns=-1},
ylabel={\#time (microseconds)},
symbolic x coords={Fetch time,Dispatch time,Execution time},
xtick=data,
nodes near coords, nodes near coords align={vertical}, ]
\addplot coordinates {(Fetch time,3884.733) (Dispatch time,8208.6) (Execution time,20109)};
\addplot coordinates {(Fetch time,4543.067) (Dispatch time,394.2) (Execution time,12018.133)};
\legend{Conceptum,Inertia} \end{axis} \end{tikzpicture}%
\end{minipage}%
\end{minipage}
\par\end{center}

\section{Conclusion}

\subsection{Analysis}

Based on the data above, we calculated the result that Inertia - the
register-based virtual machine - spends 20.39\% less time in the overall
execution time than Conceptum - our stack-based virtual machine. Inertia
is also more efficient in instruction dispatch, largely benefitted
from its small dispatch size (See section 5.1-A,B,C,D). In fact, Inertia
spends 66.42\% less time in instruction dispatch than Conceptum, on
average. However, Inertia is still slower in the overall fetch time,
spending 23.5\% more time on average in fetching operands than Conceptum
does. All of this data is consistent with Davis et. al's equation,
as presented in Section 4.1 \cite{Davis:2003:CVR:858570.858575}.
It is also imperative to note, that Inertia still has the advantage
of a smaller bytecode size.

The data presented in Figures 5.1-A,B,C,D also proved Davis et. al's
original equation estimating the performance differences of VSMs and
VRMs, as described in section 4.1, to be correct. Register-based virtual
machines spend significantly less amounts of dispatch than stack-machines
do. By analyzing the data above, one could easily see that at average
a register-based virtual machine would only execute about half the
dispatches a stack-based virtual machine would do. As one could see
in the benchmark results presented in Figure 5.2-A,B,C,D, having fewer
dispatches had won Inertia a huge advantage.

Looking specifically at the tests, an observation can be made that
a stack-based virtual machine still outperforms a register-based virtual
machine on a few occasions. Based on our test results, stack-based
virtual machines typically perform better on benchmarks featuring
a high amount of arithmetic operations. The reason behind the stack
machine's occasional good performance might be its advantage in the
fewer amount of fetches performed per dispatch. Since most arithmetic
operations in Conceptum do not require any operands at all, performing
such operations would certainly be faster than performing the same
operations in Inertia, as Inertia, as a register machine, would still
need operands to complete arithmetic operations. In contrast to the
stack-based virtual machine's performance, the register-based virtual
machine performed much better on recursions and memory operations.

After looking at the data above, one could conclude that a well-crafted
register-based virtual machine is faster both overall in execution
time and in instruction dispatch than a typical stack-based virtual
machine. Although the stack-based virtual machine had displayed its
advantages in operands fetch time, the advantages of the register-based
virtual machine in the dispatch phase affected the final results even
more. Because of the advantages in speed and its small bytecode size,
a register-based virtual machine can be best applied on systems where
both computational and storage facilities are limited (e.g. embedded
systems). In contrast, the stack machine, still benefiting from its
advantage of the clean logic (as there's no need to specify a memory
address) and rapidness in fetching instructions, can be applied to
mainstream languages requiring a balance between performance and usability.

\bibliographystyle{plain}
\bibliography{addon,MomentofInertia,Conceptual-Inertia}

\begin{thebibliography}{10}

\bibitem{mcglashan:1999}
A.~Bower B.~McGlashan.
\newblock The interpreter is dead (slow). isn't it?
\newblock {\em OOPSLA'99 Workshop: Simplicity, Performance and Portability in
  Virtual Machine design}, 1999.

\bibitem{NewBaker:2016}
T.~P. Baker.
\newblock Benchmarks and time metrics, 2016.

\bibitem{Davis:2003:CVR:858570.858575}
Brian Davis, Andrew Beatty, Kevin Casey, David Gregg, and John Waldron.
\newblock The case for virtual register machines.
\newblock In {\em Proceedings of the 2003 Workshop on Interpreters, Virtual
  Machines and Emulators}, IVME '03, pages 41--49, New York, NY, USA, 2003.
  ACM.

\bibitem{ertl:1996}
M.~A. Ertl.
\newblock Implementation of stack-based languages on register machines.
\newblock {\em PhD thesis, Technische Universit¨at Wien, Austria}, 1996.

\bibitem{Ertl:1995:SCI:207110.207165}
M.~Anton Ertl.
\newblock Stack caching for interpreters.
\newblock In {\em Proceedings of the ACM SIGPLAN 1995 Conference on Programming
  Language Design and Implementation}, PLDI '95, pages 315--327, New York, NY,
  USA, 1995. ACM.

\bibitem{fagerholm:2005}
Fabian Fagerholm.
\newblock Perl 6 and the parrot virtual machine.
\newblock 2005.

\bibitem{fang:201608}
R.~Fang.
\newblock Conceptual-inertia/conceptum: Optimized release.
\newblock DOI: 10.5281/zenodo.160870, oct 2016.

\bibitem{fang:2016160466}
R.~Fang.
\newblock frjalex/memman: memman v0.1beta.
\newblock DOI: 10.5281/zenodo.160466, oct 2016.

\bibitem{lerusalimschy:2005}
Roberto Ierusalimschy and Luiz~Henrique De~Figueiredo.
\newblock The implementation of lua 5.0.
\newblock 2005.

\bibitem{Koslowski:2013}
Jurgen Koslowski.
\newblock Deterministic single-state 2pdas are turing-complete.
\newblock {\em Technischer Bericht}, 2013.

\bibitem{liu:2016888}
S.~Liu.
\newblock Conceptual-inertia/moment-of-inertia: Iitsec 2016 release.
\newblock DOI: 10.5281/zenodo.161471., oct 2016.

\bibitem{Myers:1977:CAS:859402.859403}
Glenford~J. Myers.
\newblock The case against stack-oriented instruction sets.
\newblock {\em SIGARCH Comput. Archit. News}, 6(3):7--10, August 1977.

\bibitem{Schulthess:1977:RCA:859412.859416}
Peter~U. Schulthess and Eduard~P. Mumprecht.
\newblock Reply to the case against stack-oriented instruction sets.
\newblock {\em SIGARCH Comput. Archit. News}, 6(5):24--27, December 1977.

\bibitem{Shi:2005:VMS:1064979.1065001}
Yunhe Shi, David Gregg, Andrew Beatty, and M.~Anton Ertl.
\newblock Virtual machine showdown: Stack versus registers.
\newblock In {\em Proceedings of the 1st ACM/USENIX International Conference on
  Virtual Execution Environments}, VEE '05, pages 153--163, New York, NY, USA,
  2005. ACM.

\bibitem{NewGnuClockTime}
GNU's~Not UNIX.
\newblock The gnu c library reference manual, date and time.

\bibitem{winterbottom:1997}
P.~Winterbottom and R.~Pike.
\newblock The design of the inferno virtual machine.
\newblock 1997.

\bibitem{NewZakeski}
Mathew Zaleski.
\newblock Yeti: a greadully extensible trace interpreter.
\newblock 2008.

\end{thebibliography}

\part*{\newpage{}}

\part*{Additional Materials}

\section*{I. The Conceptum virtual machine's bytecode format and instruction
set}

The complete instruction set for Conceptum is listed below:

\begin{lstlisting}

iconst <integer>  | store an integer constant into the local stack
fconst <float>    | store a floating point number into the local stack
cconst <char>     | store a character into the local stack
bconst <boolean>  | store a boolean value, 0 or 1

iadd              | integer addition
imul              | integer multiplication
idiv              | integer division
fadd              | floating point addition
fmul              | floating point multiplication
fdiv              | floating point division

if                | boolean if clause, equivalent to (~p) V q
ne                | negation
and               | boolean AND
or                | boolean OR
xor               | boolean XOR
dup               | duplicate the value on top of the local stack
swap              | pop 1st and 2nd value out of local stack and swap them
inc               | increase the value on top of the stack
dec               | decrease the value on top of the stack

pop               | pop 1 value out of the local stack

gload             | Pop a value from the global stack and store locally
gstore            | Pop a value from local stack and store globally

procedure main    | the starting procedure of a program
                  | must be declared on the first line of file
procedure <name>  | define a procedure call
call              | call a procedure
ret               | end of a procedure definition
ter               | return in midway to the parent procedure
goto
                  | goto another line in the same procedure
                  | (0 as procedure's first line)
if_icmple <line number>
                  | if value on top of stack evaluates to false, goto line number
\end{lstlisting}

\section*{II. The Inertia virtual machine's bytecode format and instruction
set}

Inertia's bytecode, similar to Conceptum, is written in Polish notation,
and then compiled by the ``Moment of Inertia'' compiler to binary
format. The complete instruction set of Inertia is listed below:

\begin{lstlisting}

Inertia's Bytecode format:

 - WRITTEN IN POLISH NOTATION.
 - INSTRUCTION TYPE FIRST, THREE PARAMETER (OPTIONAL) AFTER,
 - SEPARATED BY SPACE;
 - INSTRUCTIONS ARE SEPARATED BY LINES

INSTRUCTIONS:
add        |add value and assign to register
div        |division value and assign to register
mul        |multiple value and assign
ltn        |return true if value second less than third argument
eql        |return true if value second equal to third argument
and        |bitwise and
not        |bitwise not
or         |bitwise or
inc        |increase value by 1
dec        |decrease value by 1
print      |print argument
load       |assign second argument to 1
goto       |goto statement
if         |has the same meaning as
           | to if_icmple in Conceptum's instruction

return     |return from a procedure
call       |call a procedure

PARAMETER TYPES:
@ memory address
R register address
# number
P goto label

GOTO LABEL:
The goto labels are labeled by numbers, starting at 1.
Goto labels used in argument format:
P <LABEL>
Example:
<LABEL>: <code>
\end{lstlisting}

The binary format of Inertia's bytecode, assembled by the assmbler
'Moment of Inertia', can then be parsed by the Inertia virtual machine
itself. The binary format is organized as follows:

\begin{lstlisting}
ADD 0x0            // Addition
DIV 0x1            // Division
MUL 0x2            // Multiplication
LTN 0x3            // Less Than
EQL 0x4            // Equal To
AND 0x5            // Bitwise AND
INERTIA_NOT 0x6    // Bitwise NEGATION
OR 0x7             //  Bitwise OR
INC 0x8            // Increase by 1
DEC 0x9            // Decrease by 1
PRINT 0xA          // Print to stdout
LOAD 0xB           // Load value 
GOTO 0xC           // goto 
IF 0xD    //if par1 is false, goto line specified by par2
RETURN 0xE //return
CALL 0xF  // call function

Hexadecimal format
0 :memory,
1 :register,
2 :const

instruction :first 4 bits
type of input :2 bits, 2*3 = 6 bits
field for register :2 bits, 2*3 = 6bits
field for memory :16 bits, 16 + 16* 2 = 48 bits
\end{lstlisting}

\section*{III. Source code listings of the programs used in benchmark}

The four benchmark programs written seperately for Conceptum and Inertia
are listed as follows. The bytecode suffix Conceptum uses is .fng;
the bytecode suffix Inertia uses is .gnf.

\medskip{}

\noindent %
\noindent\begin{minipage}[t]{1\columnwidth}%
\noindent %
\begin{minipage}[t]{0.45\columnwidth}%

\subsection*{Fibonacci\_stack.fng}

\begin{lstlisting}
procedure main
iconst 0
gstore
iconst 1
gstore
iconst 0
dup
iconst 1000
swap
ilt
if_icmple 20
inc
gload
gload
swap
dup
gstore
iadd
gstore
goto 5
gload
print
ret
\end{lstlisting}
\end{minipage}\hfill{}%
\begin{minipage}[t]{0.45\columnwidth}%

\subsection*{Fibonacci\_register.gnf}

\begin{lstlisting}
load R1 #0
load R2 #500
load @0 #0
load @1 #1
1:
ltn R0 R1 R2
if R0 P2
add @0 @0 @1
add @1 @0 @1
inc R1
goto P1
2:
print @1
return
\end{lstlisting}
\end{minipage}%
\end{minipage}\medskip{}
\noindent\begin{minipage}[t]{1\columnwidth}%
\begin{minipage}[t]{0.45\columnwidth}%

\subsection*{AddictiveAddition\_stack.fng}

\begin{lstlisting}
procedure main
iconst 0
dup
iconst 5000000
igt
if_icmple 7
inc
goto 1
ret
\end{lstlisting}
\end{minipage}\hfill{}%
\begin{minipage}[t]{0.45\columnwidth}%

\subsection*{AddictiveAddition\_register.gnf}

\begin{lstlisting}
load R1 #0
load R2 #5000000
1:
ltn R0 R1 R2
if r0 P2
inc R1
goto P1
2:
return
\end{lstlisting}
\end{minipage}%
\end{minipage}

\noindent \medskip{}
\noindent\begin{minipage}[t]{1\columnwidth}%
\begin{minipage}[t]{0.45\columnwidth}%

\subsection*{ExhaustiveCollatz\_stack.fng}

\begin{lstlisting}
procedure main
iconst 1
dup
iconst 20000
swap
ilt
if_icmple 31
dup
dup
iconst 1
ieq
ne
if_icmple 28
dup
iconst 1
and
dup
if_icmple 22
swap
iconst 3
imul
inc
swap
ne
if_icmple 27
iconst 2
swap
idiv
goto 7
pop
inc
goto 1
ret 
\end{lstlisting}
\end{minipage}\hfill{}%
\begin{minipage}[t]{0.45\columnwidth}%

\subsection*{ExhaustiveCollatz\_register.gnf}

\begin{lstlisting}
load R1 #1
1:
ltn R0 R1 #20000
if R0 P2
load @1 R1
3:
eql R2 @1 #1
not R2 R2
if R2 P4
and R3 @1 #1
eql R3 R3 #1
if R3 P5
mul @1 @1 #3
inc @1
5:
not R3 R3
if R3 P6
div @1 @1 #2
6:
goto P3
4:
inc R1
goto P1
2:
return
\end{lstlisting}
\end{minipage}%
\end{minipage}\medskip{}
\noindent\begin{minipage}[t]{1\columnwidth}%
\begin{minipage}[t]{0.45\columnwidth}%

\subsection*{Recursion\_stack.fng}

\begin{lstlisting}
procedure main
iconst 1000
gstore
call rec
ret
procedure rec
gload
dup
iconst 0
ieq
if_icmple 6
ter
dec
gstore
call rec
ret 
\end{lstlisting}
\end{minipage}\hfill{}%
\begin{minipage}[t]{0.45\columnwidth}%

\subsection*{Recursion\_register.gnf}

\begin{lstlisting}
load R1 #0
load R2 #1000
call P1
print R2
return
1:
2:
ltn R0 R1 R2
if R0 P3
dec R2
call P2
3:
return
\end{lstlisting}
\end{minipage}%
\end{minipage}

\section*{IV. Getting the source code of Conceptum and Inertia}

Conceptum is primarily the work of Ruijie Fang. Conceptum's source
code is opensourced under the GNU General Public License v3.0. The
complete source code for the Conceptum virtual machine can be found
at the URL:
\begin{doublespace}
\begin{center}
\url{https://github.com/Conceptual-Inertia/Conceptum}.
\par\end{center}
\end{doublespace}

\noindent Inertia is mainly the work of Siqi Liu. Inertia's source
code is also opensourced under the GNU General Public License v3.0.
The complete source code for the Inertia virtual machine can be found
at the URL:
\begin{doublespace}
\begin{center}
\url{https://github.com/Conceptual-Inertia/Inertia}.
\par\end{center}
\end{doublespace}

\noindent All of the test codes presented in this paper can be accessed
online via the following URLs:
\begin{doublespace}
\begin{center}
\url{https://github.com/Conceptual-Inertia/testcodes}, and \url{https://github.com/Conceptual-Inertia/conceptum_testcodes}.
\par\end{center}
\end{doublespace}

\end{document}